\documentclass[a4paper,12pt]{article}
\usepackage[T2A]{fontenc}
\usepackage[utf8]{inputenc}   
\usepackage[russian,english]{babel}

\usepackage{verbatim} 	
\usepackage{amsmath, amsthm}
\usepackage{graphicx}
\usepackage{array} 	

\usepackage{rotating}	

\usepackage{hyperref}	

\begin{document}

\title{Modified exponential I(U) dependence and optical efficiency 
of AlGaAs SCH lasers in computer modeling with Synopsys TCAD.}

\author{Zbigniew Koziol, and Sergey I. Matyukhin\\
Orel State Technical University, 29 Naugorskoye Shosse,\\
Orel, 302020, Russia}

\maketitle

\begin{abstract}
Optical and electrical characteristics of AlGaAs lasers with separate confinement heterostructures
are modeled by using Synopsys's Sentaurus TCAD, and open source software. 
We propose a modified exponential $I-V$ dependence to describe electrical properties.
A simple analytical, phenomenological model is found to describe optical efficiency, $\eta$, with a high accuracy, 
by using two parameters only. A link is shown between differential electrical resistivity $r=dU/dI$ 
just above the lasing offset voltage, and the functional $\eta(U)$ dependence.


\end{abstract}


\section{Introduction}
\label{Introduction}

Alferov \cite{Alferov}, et al., proposed creating semiconductor-based lasers
comprising the use of a geometrically-narrow active recombination region 
where photon generation occurs, 
with waveguides around improving the gain to loss ratio
(separate confinement heterostructures; SCH). That idea dominated largely 
the field of optoelectronics development
in the past years. Due to the relative simplicity and perfection of technology,
solid solutions of $Al_x Ga_{1-x}As$ are commonly used as wide-gap semiconductors in SCH lasers.

Reaching the threshold current density of these lasers less than $1 kA/cm^2$ at room 
temperature has opened up prospects for their practical application and served 
as a turning point in their production. Now, they are mostly used for pumping solid state $Nd:YAG$ lasers, 
either for high-power metallurgical processes or, already, in first field experiments 
as a highly directional source of energy in weapons interceptors.
Further progress in that direction is associated with optimizing the design of laser diodes and, in particular, 
in improving their optical efficiency as well finding methods of removing excess heating released.

In our earlier works we first were able to find agreement between our calculations of quantum 
well energy states and the lasing wavelength observed experimentally \cite{Koziol}. 
Next \cite{Koziol2}, we have shown how to considerably improve their electrical and optical 
parameters by finding the most optimal QW width and waveguides widths, and type and level of doping \cite{Koziol3}.
We compared computed properties with these of lasers produced by Polyus research 
institute in Moscow \cite{Andrejev}, \cite{Andrejev_2}.
By changing the waveguide profile through introducing a gradual
change of Al concentration, as well variable doping profiles, we were able to decrease 
significantly the lasing threshold current, increase the slope of optical power versus current, 
and increase optical efficiency. 

We have shown also \cite{NGC} that the lasing action may not occur at certain widths
or depths of Quantum Well (QW), and the threshold current as a function of these parameters may have 
discontinuities that occur when the most upper quantum well energy values are very close to either 
conduction band or valence band energy offsets. These effects are more pronounced at low temperatures, 
and may be observed also, at certain conditions, in temperature dependence of lasing threshold current 
as well.

One of the fundamental laser characteristics is their optical efficiency, $\eta$, the ratio of optical power generated, $L$,
to electrical power supplied, $\eta=L/(U\cdot I)$, as well dependence of $\eta$ on current or voltage.
We propose here a simple analytical, phenomenological model for description of $I-V$ characteristics 
near and above lasing offset voltage $U_0$, and we show that that model may be used for description 
of $\eta(U)$ with a high accuracy. 
At the same time we obtain a link between differential electrical resistivity $r=dU/dI$ just above 
$U_0$, with the functional $\eta(U)$ dependence, where $r$ is also an important experimental 
characteristic of a laser device. 

For simulations, we use Sentaurus TCAD from Synopsys \cite{tcad}, which is an advanced commercial 
computational environment, a collection of tools for performing modeling of electronic devices. 


\section{Lasers structure and calibration of modeling.}
\label{Lasers_structure}

We model a laser with $1000  \mu m$ cavity length and $100  \mu m$
width, with doping/Al-content as described in Table \ref{table_1}.

Synopsys's Sentaurus TCAD is a flexible set of tools used for modeling 
a broad range of technological and physical 
processes in the world of microelectronics phenomena. It can be run on Windows and Linux OS. 
Linux, once mastered, offers more ways of an efficient solving 
of problems by providing a large set of open source tools and ergonomic environment for their use, making
it our preferred operating system. We find it convenient, for instance, to use 
Perl\footnote{Perl stands for \emph{Practical Extraction and 
Report Language}; http://www.perl.org} scripting language for control 
of batch processing and changing parameters of calculations as well for manipulation on text data files,
and Tcl\footnote{\emph{Tool Command Language}; http://www.tcl.tk} for manipulating (extracting) 
spacial data from binary $TDR$ files. A detailed description, with examples of scripts, is available 
on our laboratory web site\footnote{http://www.ostu.ru/units/ltd/zbigniew/synopsys.php}.

The results for $I_{th}$ and $S=dL/dI$, in this paper, are all shown normalized by $I_{th}^0$ and $S_0$, 
respectively, which are the values of $I_{th}$ and $S$ computed for the reference laser 
described in Table \ref{table_1}.

We neglect here the effect of contact resistance, $R_x$, 
by not including buffer and substrate layers and contacts into calculations
(compare with structure described in Table \ref{table_1}). 
An estimate, based on geometric dimensions of substrate layers and their microscopic
parameters (doping concentration, carrier mobility) gives us a value of $R_x$ 
of the order of $13 m\Omega$. At lasing threshold current of $0.1 A$, 
that small resistance will cause a difference between computed by us 
lasing offset voltage $U_0$ and that measured one by about $1 mV$ only. 
We still will however have a noticeable contribution from $R_x$ to differential
resistance $dU/dI$.


\section{Methods of data analysis.}
\label{Data_analysis}

\subsection{Threshold current and $L(I)$ dependence.}
\label{L(I)}

The most accurate way of finding $I_{th}$ is by extrapolating the linear part of $L(I)$ 
to $L=0$ just after the current larger than $I_{th}$. We used a set of gnuplot and perl scripts for that
that could be run semi-automatically, very effectively, on a large collection of datasets.
One should only take care that the data range for fiting is properly chosen,
since $L(I)$ is a linear function in a certain range of $I$ values only. The choice of that
range may affect accuracy of data analysis. 

\subsection{$U_0$ from fiting $U(I)$ dependence}
\label{From_U(I)}

An exponential $U(I)$ dependence is found to work well at voltages which are well below
the lasing offset voltage $U_0$. Near the lasing threshold, we observe 
a strong departure from that dependence, 
and, in particular, for many data curves a clear kink in $U(I)$ is observed at $U_0$. 
We find that a modified exponential dependence describes the data very well:

\begin{equation}\label{exponential_6_parameters}
\begin{array}{ll}
	I(U) = I_{th} \cdot exp(A\cdot (U-U_0) + B \cdot (U-U_0)^2),~~~ for ~ U< U_0\\
	I(U) = I_{th} \cdot exp(C\cdot (U-U_0) + D \cdot (U-U_0)^2),~~~ for ~ U> U_0
\end{array}
\end{equation}

where $I_{th}$, $U_0$, as well $A$, $B$, $C$, and $D$ are certain fiting parameters. Equation  
\ref{exponential_6_parameters} offers a convenient interpretation of physical meaning of it's 
parameters $I_{th}$ and $U_0$: $I(U_0) = I_{th}$. 

\subsection{Differential resistance}
\label{resistance}

\begin{figure}[t]
\begin{center}
      \resizebox{150mm}{!}{\includegraphics{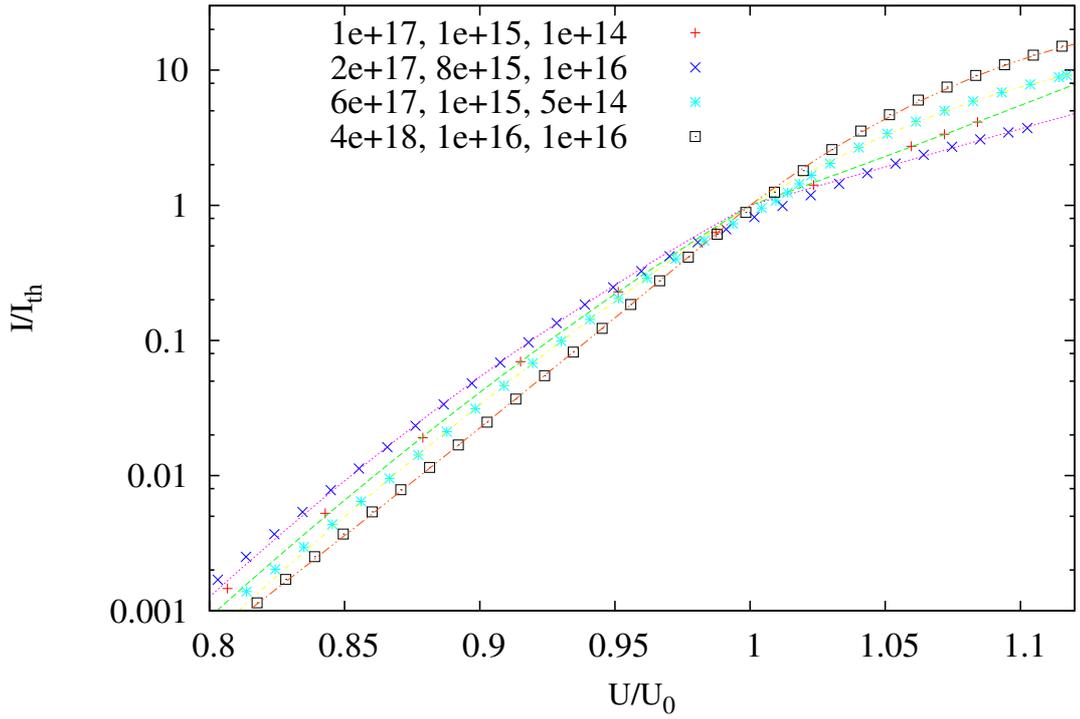}}
      \caption{Examples of typical $I-V$ characteristics for a few combination (as described
in the Figure) of doping concentrations (n- and p-emitters concentration first, followed by waveguides
and active region concentrations). We show the narrow region near the lasing threshold, only.
The curves are computed by using fiting parameters ($I_{th}, A, B, C, D, U_0$) of equation \ref{exponential_6_parameters},
and after that voltage and current are normalized by $U_0$ and $I_{th}$, respectively. 
}
      \label{doping21a}
\end{center}
\end{figure}

The above function (Eq. \ref{exponential_6_parameters}) is continuous at $U_0$, as it obviously should, 
but it's derivative is usually not. Figure \ref{doping21a} shows a few typical examples of $I(U)$ datasets.
The lines were computed analytically by using Eq. \ref{exponential_6_parameters}, 
after finding all parameters with the least-squares method.

Since (\ref{exponential_6_parameters}) may have a discontinuous derivative,
using it to find out differential resistance at $U_0$ is ambiguous. 
From Eq. (\ref{exponential_6_parameters}), at $U=U_0$, we will have $dU/dI=\frac{1}{I_{th} \cdot A}$
on the side $U<U_0$ and $dU/dI=\frac{1}{I_{th} \cdot C}$ on the side $U>U_0$. Hence, the parameter $C$
may be interpreted in terms of differential resistivity just above $U_0$:

\begin{equation}\label{exponential_6_parameters_5}
\begin{array}{ll}
r = \frac {1}{C\cdot I_{th}}
\end{array}
\end{equation}

We find from data analysis, for instance for the third dataset in Figure \ref{doping21a},
that $dU/dI \approx 50 m\Omega$, which, together with estimated contact resistance $R_x=13 m\Omega$ gives good
qualitative agreement with the differential resistance expected for real lasers, 
where it is in the range of about $50-80 m\Omega$ (\cite{Andrejev} and \cite{Andrejev_2}).


\subsection{Doping dependencies}
\label{doping}

\begin{figure}[t]
\begin{center}
      \resizebox{150mm}{!}{\includegraphics{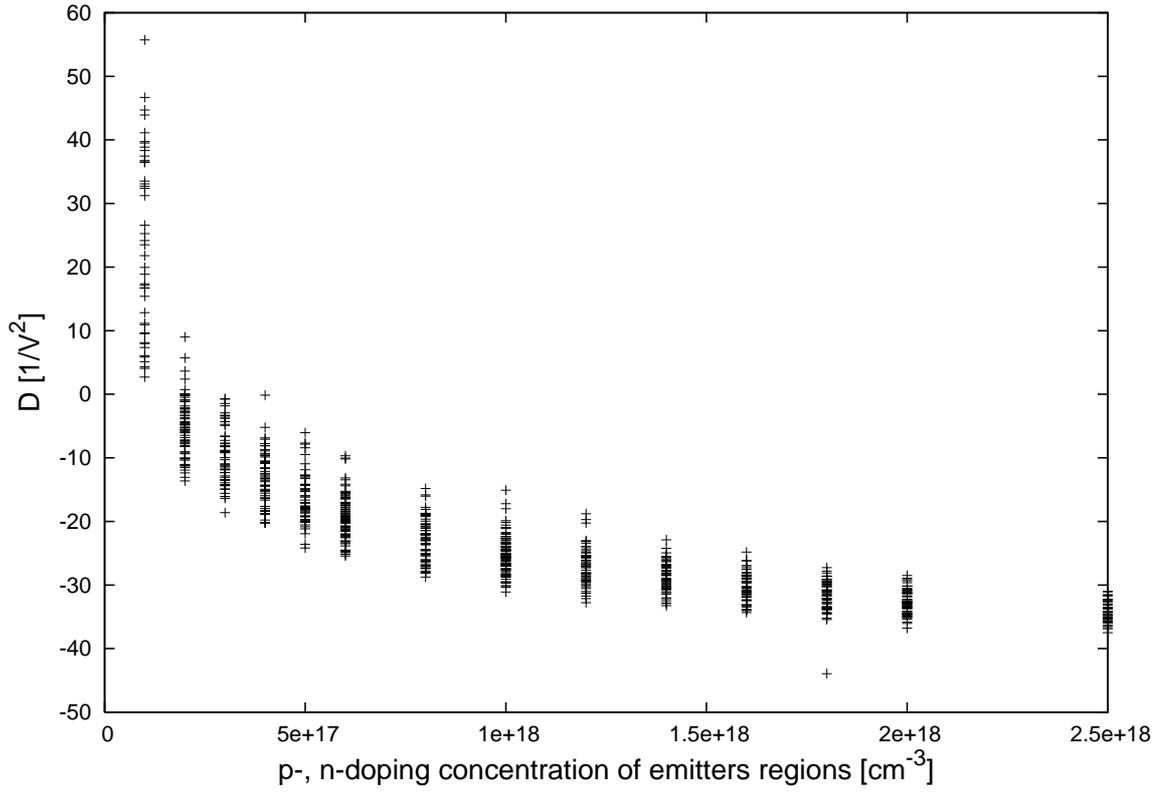}}
      \caption{Dependence of $D$ on n-, and p-emitters doping concentration,
for a broad range of doping concentrations in other regions: 
n-type concentration in active region changes between $10^{14}$ and $10^{16} cm^{-3}$,
while n-type concentration in waveguide regions between $10^{15}$ and $10^{16} cm^{-3}$.
}
      \label{doping28}
\end{center}
\end{figure}

Figure \ref{doping28} shows the dependence of parameter $D$ in Eq. \ref{exponential_6_parameters}
on n-, and p-emitters doping concentration, for a very broad range of doping concentrations in other regions 
(this is "N-N" type of doping; see description of Table \ref{table_1}). 
Due to large scatter of the parameters obtained by the least-squares fiting, we do not distinguish between 
datapoints that were obtained for various doping concentrations in waveguides or in active region: 
the dominant factor on values of $C$ or $D$ parameters is doping concentration in emitter regions.

\begin{figure}[t]
\begin{center}
      \resizebox{150mm}{!}{\includegraphics{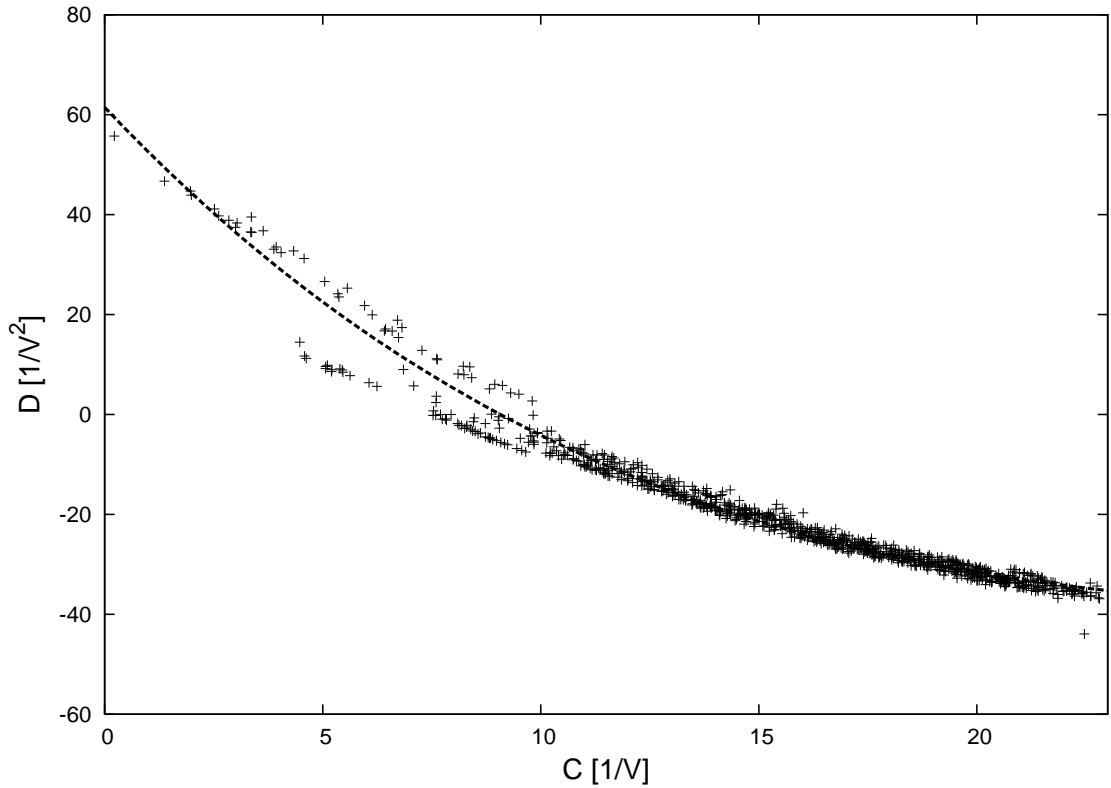}}
      \caption{Correlation between parameters $D$ and $C$
		for a broad range of doping concentrations in all regions: 
		n-type concentration in active region changes between $10^{14}$ and $10^{16} cm^{-3}$,
		n-type concentration in waveguide regions is between $10^{15}$ and $10^{16} cm^{-3}$,
		and in emitter regions it changes between $10^{17}$ and $2.5 \cdot 10^{18} cm^{-3}$.
		The line is described in the text.
}
      \label{doping30}
\end{center}
\end{figure}

We observe also that a correlation exists between values of $C$ and $D$ parameters, 
as illustrated in Figure \ref{doping30}. The line in Figure \ref{doping30} was obtained 
by using the least-squares fiting method to 
all the data points displayed there, with the following simple function:

\begin{equation}\label{parameters_11}
\begin{array}{ll}
D=-40.073 + 8.6 \cdot 10^{-5} \cdot (41.4 -C)^{3.75}
\end{array}
\end{equation}

It is convenient to rewrite Equation \ref{exponential_6_parameters} in dimensionless variables. In case of 
$U> U_0$ we have then:

\begin{equation}\label{dimensionless}
\begin{array}{ll}
	i(u) = exp\left(\frac{1}{\alpha} \cdot (u-1) \cdot \left[ 1 + \beta \cdot U_0^2 \cdot (u-1) \right] \right),
\end{array}
\end{equation}

where we defined: $i(u)=I(U)/I_{th}$ and $u=U/U_0$, $\alpha= r \cdot I_{th} / U_0$, $\beta=U_0 \cdot D/C$, 
and we used also Eq. \ref{exponential_6_parameters_5}.

Let us estimate the range of reasonable values of $\beta$ parameter.

The function \ref{parameters_11} would 
give the ratio $D/C \rightarrow +\infty$ for $C$ decreasing to $0$ (which corresponds to decreasing
doping concentration in emitter regions to $0$). That function will pass through $0$ at values of 
$C \approx 8.9$, which corresponds to doping in emitters of around $2\cdot 10^{17} cm^{-3}$, 
will have minimum of value  $\approx -1.59$ at concentrations corresponding 
to $\approx 2\cdot 10^{18} cm^{-3}$, and will increase to $-1$ at larger doping concentrations.
The practical range of interest in our case is not at the lowest doping concentrations in emitters, 
since than other laser parameters deteriorate. We are left with $D/C$ values that are important to us
in the range between $\approx -1.6$ and $0$.

Hence, the corresponding range of $\beta$ values that is of our interest is between $\approx -1$ and $0$.


\section{Optical efficiency}
\label{effciency}

\subsection{A simplified approach}
\label{simplified}

It is tempting to try a simplified version of \ref{dimensionless}, when the expression under exponent is $<< 1$.
We have in that case the following approximation on optical efficiency:

\begin{equation}\label{exponential_6_parameters_9}
\begin{array}{ll}
\eta(u) = \frac {S}{U_0 \cdot u} \cdot 
				\frac{
					 (u-1) + \beta \cdot (u-1)^2 
				}
				{
					\alpha + \left[ (u-1) + \beta \cdot (u-1)^2 \right]
				}
\end{array}
\end{equation}

\begin{figure}[t]
\begin{center}
      \resizebox{150mm}{!}{\includegraphics{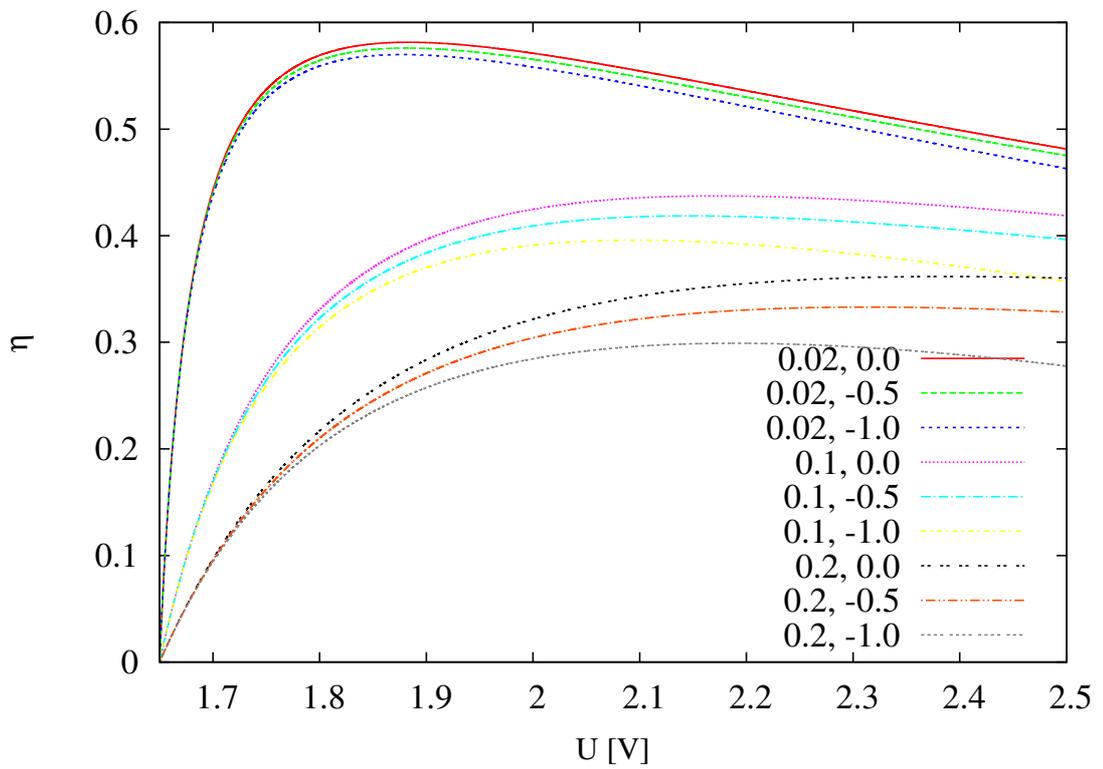}}
      \caption{Optical efficiency as a function of voltage, computed by using Equation \ref{exponential_6_parameters_9},
		for a few values of parameters $\alpha$ and $\beta$, as shown in the Figure. 
		It has been assumed that $S=1.25 W/A$ and $U_0=1.65 V$.
}
      \label{eta_02}
\end{center}
\end{figure}

Figure \ref{eta_02} shows a few example curves computed with Equation \ref{exponential_6_parameters_9}.
The accuracy of these results, if compared with real data (not shown on that figure), appears to be far from desired; Eq. \ref{exponential_6_parameters_9} may be treated
as a very rough approximation only.

\subsection{Exact result}
\label{exact}

Let us use however the full version of Equation \ref{exponential_6_parameters} (for $U>U_0$), for computing
optical efficiency. We have then:

\begin{equation}\label{exponential_6_parameters_11}
\begin{array}{ll}
\eta(u) = \frac{S}{U_0} \cdot \frac {i(u)-1}{ u \cdot i(u)},
\end{array}
\end{equation}

where $i(u)$ is given by Eq. \ref{dimensionless}.

\begin{figure}[t]
\begin{center}
      \resizebox{150mm}{!}{\includegraphics{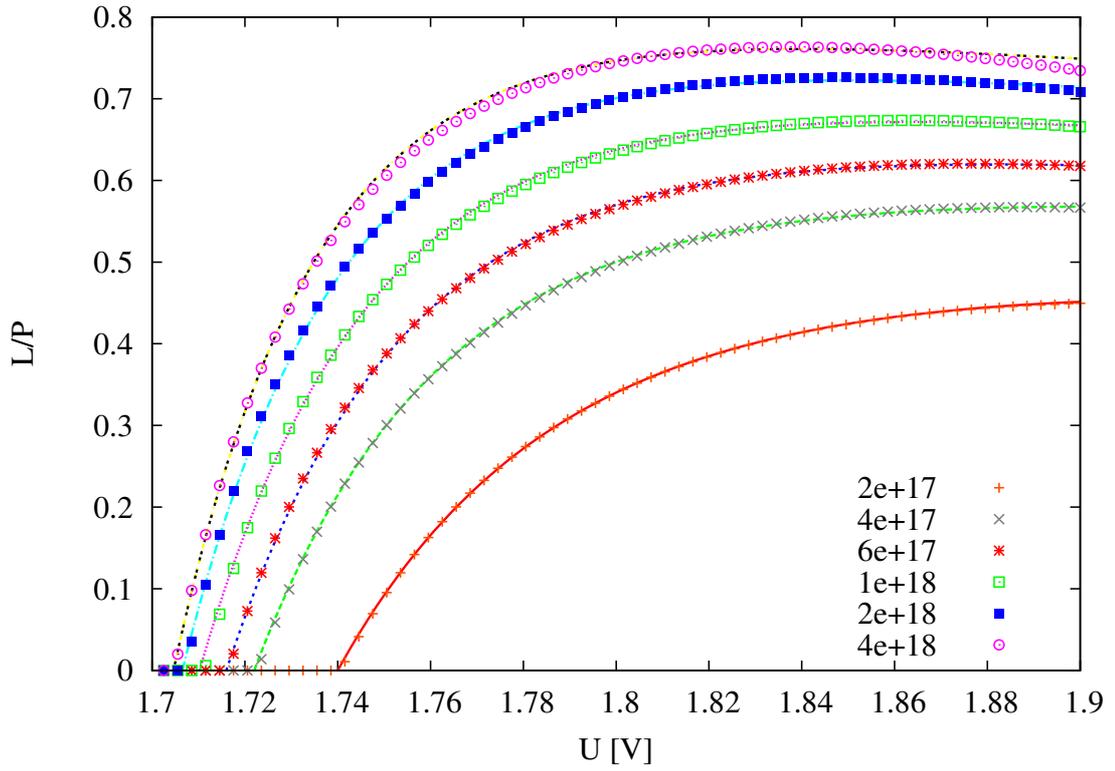}}
      \caption{The lines show optical efficiency computed with equations \ref{exponential_6_parameters_11}
		and \ref{dimensionless}, while symbols show optical efficiency directly obtained 
		from modeling data. The results are for "N-P" type of doping structure (different type of doping
		for both waveguides), for doping concentration in 
		waveguides of $1\cdot 10^{15} cm^{-3}$, in active region $5\cdot 10^{14} cm^{-3}$, 
		and in emitter regions
		as shown in the Figure.
}
      \label{eta_03}
\end{center}
\end{figure}

\begin{figure}[t]
\begin{center}
      \resizebox{150mm}{!}{\includegraphics{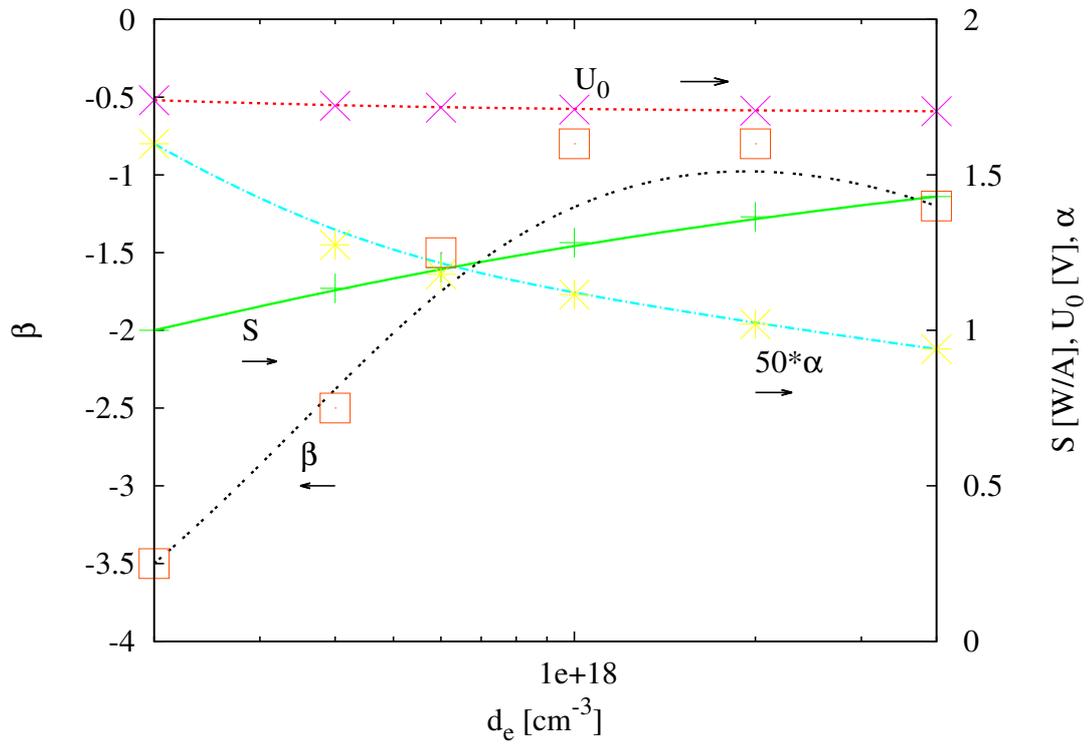}}
      \caption{Parameters $S$, $U_0$, $\alpha$ and $\beta$ as a function of doping concentration in
		emitters, that were used to draw curves in Figure \ref{eta_03}. The lines are to guide the eyes, only.
}
      \label{eta_03a}
\end{center}
\end{figure}

Figure \ref{eta_03} shows that an excellent agreement is obtained when these analytical formulas are
used for approximating optical efficiency directly computed from modeling data. The parameters used in fiting
the data are shown in Figure \ref{eta_03a}. As seen on this figure, the value of $\beta$ for some datapoints 
is lower than $-1$. This should not be considered contradictory to our estimate of the range of possible 
values of $\beta$:
our calculations of $\eta$ do not take into account the nonlinearity of $L(I)$ dependence, which may 
be large, especially for low doping concentrations in emitters, and that will effectively cause 
decrease of $\beta$ value.
We see also that the parameter $\alpha$ is too small (i.e. $(u-1)/\alpha$ is too large) 
in realistic cases to allow using the simplified 
Equation \ref{exponential_6_parameters_9}.

\clearpage

\begin{table}[t]
\caption{Structure of AlGaAs SCH laser layers used in computer modeling. Values 
of doping concentrations listed in rows 4-6 of this table are approximate only. 
In this paper we make reference to two kinds of doping concentrations in these regions, that we 
name by "N-N" and "N-P". "N-N" means that in two waveguides there is the same N-type doping concentration 
with doping type in active region of n-type as well. "N-P" means" that one waveguide is of n-type doping 
and the other one of p-type doping. Active region has then n-type doping. The "real" lasers have N-N type of doping,
however, setting "N-P" type of doping in computer modeling leads to a significant improvement of device characteristics \cite{Koziol3}.
}
\label{table_1}
\begin{center}
\begin{tabular}{|c|c|c|c|c|c|}
\hline
\bf{No}  &       \bf{Layer}&    \bf{Composition}&    \bf{Doping [$cm^{-3}$]}&    \bf{Thickness [$\mu m$]}\\
\hline
1        & n-substrate     &   n-GaAs (100)    &  $2\cdot 10^{18}$  &    350              \\
\hline
2        & n-buffer        &   n-GaAs           &  $1\cdot 10^{18}$ &    0.4              \\
\hline
3        & n-emitter       &   $Al_{0.5}Ga_{0.5}As$ & $1\cdot 10^{18}$ &    1.6              \\
\hline
4        & waveguide       &   $Al_{0.33}Ga_{0.67}As$ & none ($n \approx 10^{15}$) &  0.2                \\
\hline
5        & active region (QW) & $Al_{0.08}Ga_{0.92}As$ & none ($n \approx 10^{15}$) &  0.012                \\
\hline
6        & waveguide       &   $Al_{0.33}Ga_{0.67}As$ & none ($n \approx 10^{15}$) &  0.2                \\
\hline
7        & p-emitter       &   $Al_{0.5}Ga_{0.5}As$ &  $1\cdot 10^{18}$ &  1.6        \\
\hline
8        & contact layer   &   p-GaAs           & $4\cdot 10^{19}$ &    0.5      \\
\hline
\end{tabular}
\end{center}
\end{table}


\clearpage

\section{Summary}
\label{Summary}

Computer simulations using Sentaurus TCAD from Synopsys were used for performing modeling 
of electrical and optical characteristics of SCH lasers based on AlGaAs. 

A modified exponential $I-V$ dependence (Equations \ref{exponential_6_parameters} and \ref{dimensionless}) 
is proposed to describe electrical properties.

That simple analytical, phenomenological model is found to describe one of the most fundametal laser 
characteristics, optical efficiency, $\eta(U)$, with a high accuracy, by using two parameters only 
(except of $S=dL/dI$, $I_{th}$, and $U_0$). 
At the same time we obtain a link between differential electrical resistivity $r=dU/dI$ 
just above lasing offset voltage, with the functional $\eta(U)$ dependence. 

The proposed model is useful for both, analysis of computer modeling results
as well experimental data on real devices.

\section{Acknowledgements}
\label{Acknowledgements}
This research was carried out under the Federal Program "Research and scientific-pedagogical 
cadres of Innovative Russia" (GC number P2514).
The authors are indebted for valuable
comments and discussions to A.~A.~Marmalyuk of Research Institute "Polyus" in Moscow.



\begin{thebibliography}{9}

			\bibitem{Alferov}
			Zh.~I.~Alferov,
			\emph{The double heterostructure concept and its applications in physics, electronics, and technology},
			Rev. Mod. Phys. V.73. No.3. P.767-782 (2001).

			\bibitem{Koziol}
			S.~I.~Matyukhin, Z.~Koziol, and S.~N.~Romashyn, 
			\emph{The radiative characteristics of quantum-well active region of AlGaAs lasers 
			with separate-confinement heterostructure (SCH)}, 
			arXiv:1010.0432v1 [cond-mat.mtrl-sci] (2010)

			\bibitem{Koziol2}
			Z.~Koziol, S.~I.~Matyukhin, 
			\emph{Waveguide profiling of AlGaAs lasers with separate confinement heterostructures (SCH) for optimal optical and electrical characteristics, by using Synopsys's TCAD modeling}, 
			unpublished, (2011).

			\bibitem{Koziol3}
			Z.~Koziol, S.~I.~Matyukhin, and S.~N.~Romashyn,
			\emph{Doping effects in AlGaAs lasers with separate confinement heterostructures (SCH). Modeling optical and electrical characteristics with Sentaurus TCAD.}
			arXiv:1106.2501v1 [physics.comp-ph] (2011)

			\bibitem{Andrejev}
			A.~Yu.~Andreev, S.~A.~Zorina, A.~Yu.~Leshko, A.~V.~Lyutetskiy, A.~A.~Marmalyuk, 
			A.~V.~Murashova, T.~A.~Nalet, A.~A.~Padalitsa, N.~A.~Pikhtin, D.~R.~Sabitov, V.~A.~Simakov, 
			S.~O.~Slipchenko, K.~Yu.~Telegin, V.~V.~Shamakhov, I.~S.~Tarasov, 
			\emph{High power lasers ($\lambda$ = 808 nm) based on separate confinement 
			AlGaAs/GaAs heterostructures},
         Semiconductors, 43(4), 543-547 (2009).

			\bibitem{Andrejev_2}
			A.~V.~Andreev, A.~Y.~Leshko, A.~V.~Lyutetskiy,A.~A.~Marmalyuk, T.~A.~Nalyot, 
			A.~A.~Padalitsa, N.~A.~Pikhtin, D.~R.~Sabitov, V.~A.~Simakov, S.~O.~Slipchenko, 
			M.~A.~Khomylev, I.~S.~Tarasov, 
			\emph{High power laser diodes ($\lambda$ = 808 – 850 nm) based on asymmetric 
			separate confinement heterostructure},
			Semiconductors, 40(5), 628-632 (2006).

			\bibitem{NGC}
			Z.~Koziol, S.~I.~Matyukhin, and S.~N.~Romashyn,
			\emph{Non-monotonic Characteristics of SCH Lasers due to Discrete Nature of Energy Levels in QW},
			accepted for presentation at 
			\emph{Nano and Giga Challenges in Electronics, Photonics and Renewable Energy Symposium and Summer School}, 
			Moscow - Zelenograd, Russia, September 12-16, (2011).

			\bibitem{tcad}
			Synopsys, 
			\emph{Sentaurus Device User Guide}, 
			www.synopsys.com, (2010)



\end{thebibliography}
\end{document}